\documentclass{optica-article}
\journal{opticajournal} 

\articletype{Research Article}

\usepackage[sort&compress]{natbib}
\usepackage{booktabs}

\setcitestyle{numbers,open={[},close={]},comma}
\makeatletter
\AddToHook{begindocument/end}{%
	\renewcommand\bibnumfmt[1]{#1.}
	\renewcommand\@biblabel[1]{#1.}
	\renewcommand\NAT@biblabelnum[1]{\bibnumfmt{#1}}
	\@ifundefined{bibsep}{}{\setlength{\bibsep}{0pt}}
	\renewcommand\NAT@bibsetnum[1]{%
		\settowidth\labelwidth{\@biblabel{#1}}%
		\setlength{\leftmargin}{\labelwidth}\addtolength{\leftmargin}{\labelsep}%
		\setlength{\parsep}{0pt}%
		\setlength{\itemsep}{0pt}%
		\setlength{\parskip}{0pt}%
		\setlength{\topsep}{0pt}%
		\setlength{\partopsep}{0pt}%
		\setlength{\listparindent}{0pt}%
		\setlength{\itemindent}{0pt}%
		\ifNAT@openbib
			\addtolength{\leftmargin}{\bibindent}%
			\setlength{\itemindent}{-\bibindent}%
			\setlength{\listparindent}{\itemindent}%
			\setlength{\parsep}{0pt}%
		\fi
	}%
	\let\@bibsetup\NAT@bibsetnum
}
\makeatother
\renewcommand{\vec}[1]{\boldsymbol{#1}}

\begin{document}

\title{The 2D approximation quickly breaks down in reflection ptychography}

\author{Sander Senhorst\authormark{1,*}, 
Stefan Witte\authormark{1} and 
Wim Coene\authormark{1,2}}

\address{
\authormark{1}{Department of Imaging Physics, Delft University of Technology, Lorentzweg 1, 2628 CJ Delft, The Netherlands }\\
\authormark{2}{ASML Netherlands B.V., De Run 6501, 5504 DR Veldhoven, The Netherland }\\
}

\email{\authormark{*}s.senhorst@tudelft.nl} 

\begin{abstract*}
Ptychographic reconstructions in reflection geometries are commonly interpreted with the same two-dimensional thin-sample model used in transmission, yet the validity of this approximation has not been established. We develop a three-dimensional weak-scattering description of reflection ptychography and derive explicit thickness criteria for when a two-dimensional model remains accurate. Because the sampled axial spatial frequency range is dominated by the rotation of the Ewald sphere rather than its curvature, reflection geometries impose far stricter thin-sample conditions than transmission geometries. The allowable thickness is reduced by one to two orders of magnitude for a representative extreme ultraviolet geometry, depending on the tolerance for appearance of artifacts. Simulations verify that conventional two-dimensional reconstructions may exhibit the thickness-dependent artifacts as predicted by the theory, with particularly strong distortions near specular Bragg minima. We further show that incorporating the correct depth-dependent propagation into the forward model resolves these distortions and enables recovery of sample thickness. These results establish practical validity limits for two-dimensional reflection ptychography and identify a path toward quantitative depth-sensitive reconstructions at all geometries.
\end{abstract*}

\section{Introduction}

High-resolution imaging of surface structures enables advancement in many areas of research and technology, such as the imaging of patterned nanostructures for the field of lithography. As lithographic functional designs are set to create samples of interest with ever smaller lateral features, stacked on increasingly taller stacks\citep{bogdanowicz2025}, ptychography in reflection geometries, specifically at extreme ultraviolet (EUV) wavelengths around 13.5 nm, has emerged as a potential imaging solution that could tackle the needs of metrology and inspection for future lithographic processing nodes.

Ptychography is a form of coherent diffractive imaging (CDI), where a coherent probe beam is scanned across a sample, while at each scanning position a diffraction pattern is recorded. If the scanning is performed with sufficient overlap in the illumination of adjacent probe beam positions, the resulting collection of diffraction patterns provides enough redundancy in the data to over-constrain the otherwise ill-posed inverse problem of reconstruction of the sample function from the data. This process, known as lensless imaging, enables the reconstruction of sample images at diffraction-limited lateral resolutions without the requirement of diffraction-limited optical components. Ptychography is of special benefit in application domains where use of diffraction-limited optics is prohibitive due to high cost or difficulty in manufacturing, such as in the domain of imaging using extreme ultraviolet wavelengths.

Since many structures of interest are placed on substrates which are optically opaque, either due to thickness or strong absorption, it is often not possible to measure sufficient scattered intensity in the transmission direction, so for such samples one must resort to imaging in a reflection-type geometry. An additional complication for EUV wavelengths is that the reflection coefficients only become significant at near-grazing incidence angles. For this reason, the angle of incidence on the sample is often chosen in the $(60^\circ, 85^\circ)$-degree range\citep{gardner2012, seaberg2014}, leading to the additional complication of diffraction-pattern distortion. The theory of ptychography was originally developed for the transmission geometry\citep{rodenburg1992}. For reflection geometries, its mathematical validity was never verified, despite many experimental demonstrations of its feasibility\citep{gardner2012, seaberg2014, odstrcil2015, zhang2016, odstrcil2016, shanblatt2016, porter2017, karl2018, tanksalvala2021, esashi2023, lu2023, shao2024, tanksalvala2024, senhorst2024}.

In many of the experimental demonstrations of reflection ptychography the reconstruction phase is converted to the height by means of the optical path difference and phase shift upon reflection\citep{lu2023, shao2024}, thus modeling the effect of depth in the sample as a constant phase shift in real space. Such an approach is based on the approximation that considers single incoming and outgoing wavevectors. The bounds of this approximation are however not clearly defined. By its nature, ptychography requires a range of incoming and outgoing wavevectors to generate diversity in the detected intensity, e.g. to create a focused illumination probe. Such a model can therefore only be completely correct if all incoming and outgoing wavevectors feel an approximately constant phase shift due to sample height, which should hold only in the limit of small detection and illumination numerical apertures. Some attempts have been made to model the effect of depth in reflection ptychography\citep{gao2025}, which models layers as incoherent contributions, requiring explicit filtering of interference terms through spatial separation, and therefore only works for large thicknesses, such as modeling the reflection from the bottom of a subtrate.

In this work, we develop a theory that can model the effect of depth with explicit bounds on the maximal permissible thickness, using the same foundational assumption of weak scattering as the original ptychographic theory. Our developed theory predicts that distortions appear in reconstructions that are performed without adequately incorporating depth effects, which we verify in simulation. Finally, we show that more accurately incorporating the effect of depth in the ptychographic model can allow reconstruction of the sample thickness using the propagation geometry alone.

\section{The 2D ptychographic model}

In 2D scanning ptychographic reconstructions assumes the separability of the field $\psi$ emerging from the sample of interest as the product of two parts: one which is independent with respect to the applied shift vector $\vec{R}_k$ at positional index $k$, and the other which is translated with respect to $\vec{R}_k$, i.e.
\begin{equation}
\label{eq:PO2d}
\psi_{k}(\vec{r}) = O(\vec{r}-\vec{R}_k)P(\vec{r}),
\end{equation}
where in this case we have taken the ``object'' function $O$ to be the shifting part, and the ``probe'' function of the illumination $P$ to be the independent part.

If the measurement is performed in the Fraunhofer regime, we detect the intensity $I_k$ of the Fourier transform of the exit field $\psi_k$, i.e:
\begin{equation}
\label{eq:farfield_intensity}
I_k(\vec{\xi}) = |\tilde{\psi}_k|^2 = |\mathcal{F}_2\left\{\psi_k\right\}|^2,
\end{equation}
where $\vec{\xi}$ is the reciprocal space coordinate measured by the detector and the tilde is used two-dimensional functions in reciprocal space, with the two-dimensional Fourier transform represented by $\mathcal{F}_2$.

Purely two-dimensional samples such as $O$ in Eq.~(\ref{eq:PO2d})  exist only in theory, yet the preceding analysis is purely two-dimensional in both the real space ($\vec{r}$) and reciprocal space ($\vec{\xi}$) coordinates.

\section{Transmission and reflection geometries}

Before continuing with derivations, it is of benefit to already define specific geometries of our interest as this may aid in the visualisations by means of example. These geometries are the transmission and reflection geometries, which we parameterize in terms of

\begin{itemize}
\item the central illumination wavevector $\vec{k}_{i,0} = k_0(\sin\theta_i, 0, \cos \theta_i)$,
\item the largest angle $\Delta \theta_i$ away from $\vec{k}_{i,0}$ such that the illumination numerical aperture $\text{NA}_i = \sin\Delta \theta_i$,
\item the central detection wavevector $\vec{k}_{d,0} = k_0(\sin \theta_d, 0, \cos \theta_d)$ and
\item the largest angle $\Delta \theta_d$ away from $\vec{k}_{d,0}$ such that the illumination numerical aperture $\text{NA}_d = \sin\Delta \theta_d$,
\end{itemize}

where $\theta_i$ and $\theta_d$ are the angles with respect to the z-axis.

Defining a transmission geometry is easily done by simply take any geometry where $\theta_i = \theta_d$. For simplicity we choose $\theta_i = \theta_d = 0$ such that the the central direction of incidence is aligned with the z-coordinate. The detected part of the exit field is then propagating in the same z-direction as the illumination part. Choosing $\theta_i = \theta_d \ne 0$ does change the requirements on the thickness, but only because the thickness $t$ is defined in non-rotated coordinates. Such an experimental geometry is not usually very practical, we disregard it in further analysis.

\begin{figure}[!htbp]
\centering
\includegraphics[width=1\linewidth]{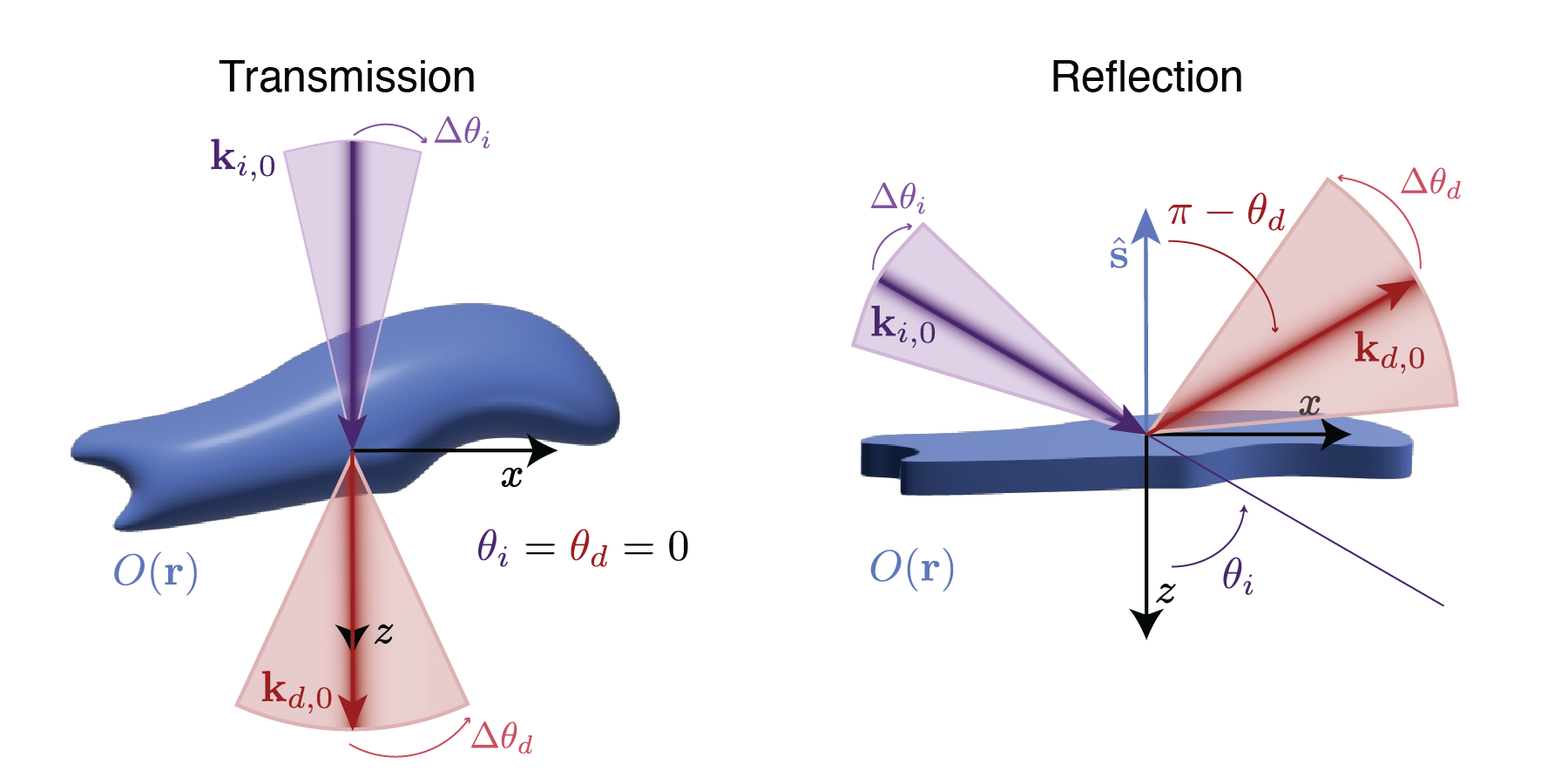}
\caption[]{The definition of the transmission geometry (left) and reflection geometry (right). The central wavevector of the illumination $\vec{k}_{i,0}$ is incident on the sample at an angle $\theta_i$ w.r.t. the $z$-axis, and the central wavevector of the detection $\vec{k}_{d,0}$ is emerging from the sample at an angle $\theta_d$ with respect to the $z$-axis. The range of the incoming wavevectors is indicated by $\Delta \theta_i$ and $\Delta \theta_d$, such that $\text{NA}_{i/d} = \sin (\Delta \theta_{i/d})$. The sample in the transmission geometry may be any function, while the definition of the reflection geometry requires presence of the normal vector $\hat{s}$ to define the direction of specular reflection.}
\label{fig:geometries}
\end{figure}

Reflection geometries are less trivial to define, since they require assumptions about the sample. What is commonly understood to be a \textit{reflection geometry} is perhaps more aptly named as a \textit{specular reflection geometry}. To define a specular reflection geometry we require the sample (or at least that area of the sample which is under investigation) to have a single planar orientation, say with normal vector $\hat{\vec{s}}$, which dominates the total back-scattered intensity. This makes it natural to place the detector such that $(\vec{k}_{d,0} - \vec{k}_{i,0}) \parallel \hat{s}$, such that the central pixel of a detector placed in the far-field is generally also the brightest, disregarding any influence due to the finite illumination NA.  If we assume we are dealing with such a sample, and we define the z-axis of our coordinate system to be along $-\hat{\vec{s}}$, then a \textit{specular reflection geometry} is defined given by $\theta_d = \pi - \theta_i$. The two different geometries have been schematically shown in Fig.~\ref{fig:geometries}.

\section{A fully three-dimensional description of scattering}

In a fully three-dimensional approach, the probe $P$ and object $O$ functions are not completely equivalent. Since the probe function corresponds to an illuminating field, under conventionally made assumptions it should be a propagating solution to the scalar Helmholtz equation. An expression for the probe function propagated to any $z$-coordinate as a function of field at $z=0$ is well-known to be given by the angular spectrum propagator \citep{goodman1996}
\begin{equation}
\label{eq:P_3d_propagate}
P(x,y,z) = \mathcal{F}^{-1}_2\left\{\exp[\pm i z k_z(\vec{\xi}_\bot)] \mathcal{F}_2\left\{ P(x, y, z=0)\right\}\right\},
\end{equation}
where $\vec{\xi}_\bot =(\xi_x, \xi_y)$ denote the transverse coordinates and where
\begin{equation}
\label{eq:kz_fn}
k_z(\vec{\xi}_\bot) = k\sqrt{1-\lambda^2\xi_x^2 - \lambda^2\xi_y^2},
\end{equation}
with wavenumber $k=2\pi/\lambda$ for wavelength $\lambda$. The sign of Eq.~(\ref{eq:P_3d_propagate}) must be chosen based on the propagation direction of the two-dimensional field; this sign is not uniquely constrained in the two-dimensional description. In a fully three-dimensional perspective, Eq.~(\ref{eq:P_3d_propagate}) may be written more concisely. Noting that the exponential factor is equivalent to the Fourier transform of a shifted impulse function of $\xi_z = \pm k_z(\vec{\xi}_\bot)$, the following expression is equivalent:
\begin{equation}
\label{eq:P_3D_ES}
P(x,y,z) = \mathcal{F}^{-1}_3\left\{\delta^\pm_{ES}(\vec{\xi}) \mathcal{F}_2\left\{P\right\}\right\},
\end{equation}
where $\mathcal{F}_3$ denotes the three-dimensional Fourier transform and where we make use of a special spherical impulse function $\delta^\pm_{ES}(\vec{\xi})$ introduced by Onural\citep{onural2006}. The subscript ${ES}$ indicates that the delta function sifts over the Ewald sphere $S$ on which time-harmonic scalar fields have to reside, defined by
\begin{equation}
\label{eq:Ewald_Sphere}
|\vec{\xi}|^2 = \xi_x^2 + \xi_y^2 + \xi_z^2 = k^2.
\end{equation}
In $\mathbb{R}^3$, $\delta_{ES}(\vec{\xi})$ is defined by the sifting property
\begin{equation}
\iiint_{\mathbb{R}^3} \delta_{ES}(\vec{\xi})f(\vec{\xi}) d^3\vec{\xi} = \iint_S f(\vec{\xi}_{ES}) d^2\vec{\xi}_{ES}.
\end{equation}
Both signs of $\delta^\pm_{ES}$ have a similar definition, only further restricting the sphere S to taking either the upper half sphere ($\xi_z\ge0$) or lower half sphere ($\xi_z<0$). Aside from some mathematical intricacies related to the Jacobian when integrating over this impulse function in three-dimensional space\citep{onural2006}, $\delta_{ES}$ may be interpreted as constraining the space of all three-dimensional functions to those with reciprocal vectors that obey the Ewald sphere condition for propagating time-harmonic fields.

The fully three-dimensional perspective favors a description in reciprocal space, since here we may make use of the sifting property of the impulse function. The entire reciprocal space is sparsely filtered out due to the 2D section of the Ewald sphere in 3D reciprocal space. We thus continue in the Fourier domain, denoting three-dimensional Fourier transforms of functions with the hat symbol (e.g. $\hat{f}$), two-dimensional transforms with the tilde (e.g. $\tilde{f}$) and one dimensional transform with the bar (e.g. $\bar{f}$).

When regarding Eq.~(\ref{eq:P_3D_ES}), we see that the sign choice for $k_z$ only allows us to describe half of the total possible solutions simultaneously. Our two-dimensional model does not intrinsically carry information on the sign of the propagation $z$-direction, which may be in either the positive or negative direction. Therefore a general field description requires two different two-dimensional probes, one of which travels in the positive $z$-direction ($P^+$) and another in the negative $z$-direction ($P^ -$):
\begin{equation}
\label{eq:Phat}
\hat{P}(\vec{\xi}) = \delta^+_{ES}(\vec{\xi})\tilde{P}^+(\vec{\xi}_{\bot})+\delta^-_{ES}(\vec{\xi})\tilde{P}^-(\vec{\xi}_{\bot}).
\end{equation}
In general, we recommend the mental model where the ``true'' probe is considered to be the fully three-dimensional function $\hat{P}$ confined to the sphere $S$, while only further restricting the allowed reciprocal vectors when the specific geometry or constraints demand it, for example by demands on the propagation direction or limited numerical aperture.

\begin{figure}[!htbp]
\centering
\includegraphics[width=1\linewidth]{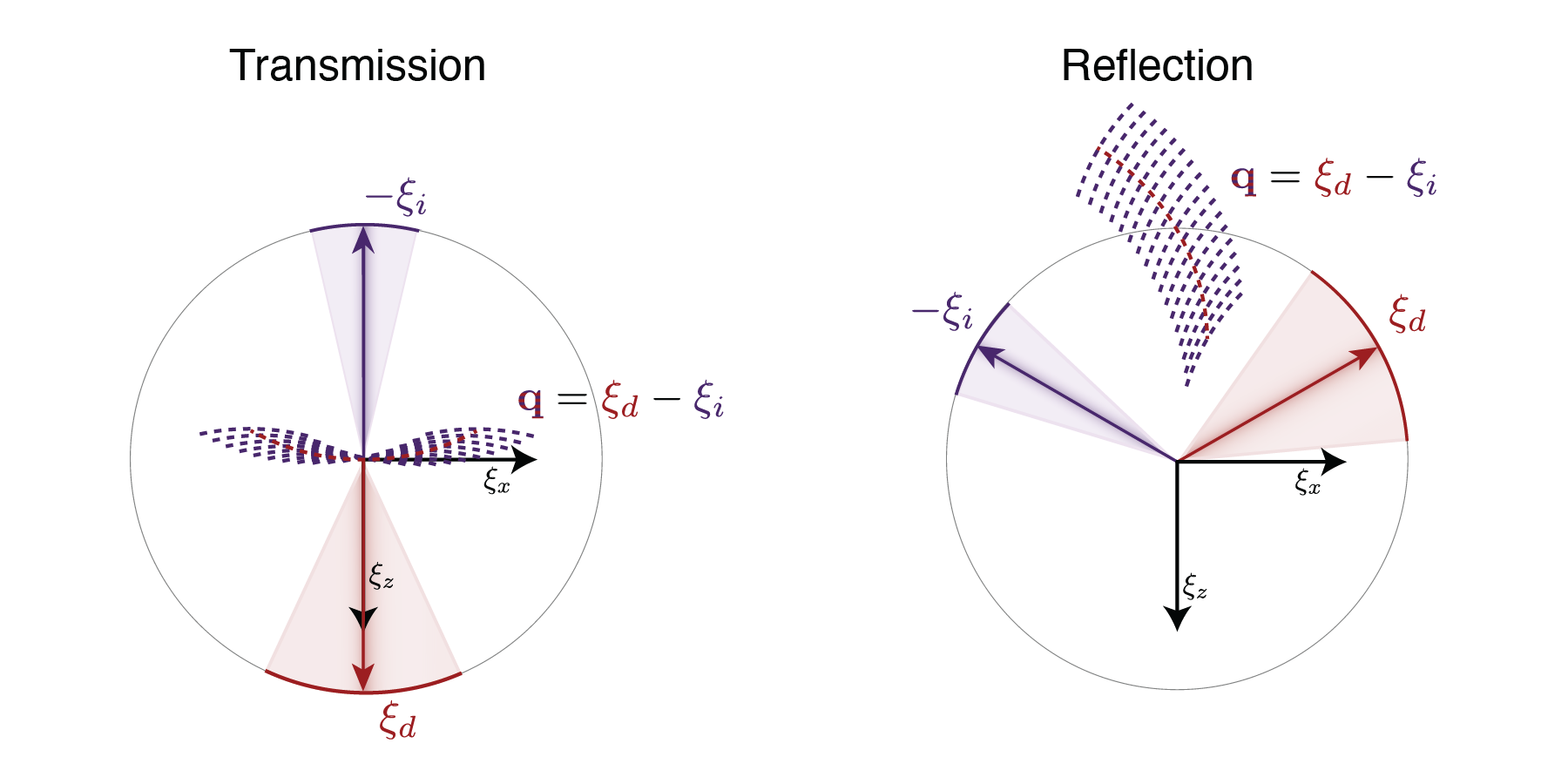}
\caption[]{The experimental geometry in the $(\xi_x,\xi_z)$ Fourier plane. The dotted lines indicate all possible vectors $\vec{\xi}_d - \vec{\xi}_i$, and thus indicate the range of $\vec{q} = \vec{\xi}_d - \vec{\xi}_i$ in Eq.~(\ref{eq:FDT_impulse}).}
\label{fig:reflection_2d_fourier}
\end{figure}

From this three-dimensional perspective on wave propagation we may attempt to compute a scattered field due to interaction with a sample of interest. If we assume the object to be weakly scattering, we may use the first Born approximation to obtain the three-dimensional near-field as a simple product between now three-dimensional fields
\begin{equation}
\label{eq:PO_3d}
\psi_k(x,y,z) = P(\vec{r})O(\vec{r}-\vec{R}_k),
\end{equation}
where now $\vec{r}$ and $\vec{R}_k$ are three-dimensional vectors, noting that standard ptychographic experiments will have $\vec{R}_k \cdot \hat{z} = 0$, although this description does not require it. Eq.~(\ref{eq:PO_3d}) is only nearly correct as it still includes those plane-wave components that correspond to modes which do not propagate to the far-field, i.e. which do not satisfy the Ewald sphere condition of Eq.~(\ref{eq:Ewald_Sphere}). In other words, $P$ satisfies the Ewald sphere condition, but $O$ is any function of three variables, so their product in real space may fill the entire reciprocal space. A correct and quite concise statement is straightforward to make in reciprocal space using $\delta_{ES}(\vec{\xi})$,
\begin{equation}
\label{eq:FDT_impulse}
\begin{aligned}
\hat{\psi}_k(\vec{\xi}_d) &=  \delta_{ES}(\vec{\xi}_d)\left\lbrack \hat{P} \circledast_3 \hat{O}_k\right\rbrack(\vec{\xi}_d)\\
&= \delta_{ES}(\vec{\xi}_d) \iiint \hat{O}_k(\vec{\xi}_d - \vec{\xi}_i)\hat{P}(\vec{\xi}_i)d^3\vec{\xi}_i,
\end{aligned}
\end{equation}
where $\circledast_3$ denotes a three-dimensional convolution and we introduce the shorthand notation
\begin{equation}
\hat{O}_k(\vec{\xi}) = \mathcal{F}_3\{O(\vec{r} - \vec{R_k})\} = \hat{O}(\vec{\xi})\exp[i \vec{\xi} \cdot \vec{R}_k].
\end{equation}
We switch here to the label $\vec{\xi}_d$ for the coordinates of the exit field since these correspond to coordinates whose sampling will be determined geometrically by the placement of the detector. Eq.~(\ref{eq:FDT_impulse}) selects only outgoing or scattered waves that reside on the Ewald sphere of the same radius $k$, which means that we are assuming an elastically scattering sample with no energy loss. Additionally, it is an extension to the Fourier diffraction theorem (FDT) \citep{wolf1969, zhou2021} first published by Wolf. The explicit coordinate relations from the original formulation, which arose by taking a plane-wave probe, are now implicit in the spherical impulse functions; furthermore, we immediately describe arbitrary incident fields. One should take special note of the scattering vector $\vec{q} = \vec{\xi}_d - \vec{\xi}_i$, which is especially relevant since it corresponds to the part of the three dimensional reciprocal space of the sample which is present in the convolution integral and which is thus ``sampled'' by the elastically scattered field. This vector may be recognizable to readers familiar with the theory of X-ray diffraction, where it often bears the name ``diffraction vector''. The effect of the convolution in Eq.~(\ref{eq:FDT_impulse}) for the transmission and reflection geometry has been schematically shown in Fig.~\ref{fig:reflection_2d_fourier}.

\section{Cases for which two-dimensional descriptions fail}

Having developed the fully three-dimensional perspective on weakly-scattering experiments, we may now use it to derive conditions where a two-dimensional approximation would be insufficient to reconstruct the ptychographic dataset. In other words, we will investigate conditions where it is possible to find two-dimensional approximating functions $O'(x,y)$ and $P'(x,y)$ which predict nearly the same exit fields as the fully 3D model and can therefore provide a valid model for reconstructing the dataset. The exit field of such a fully two-dimensional model is given by the standard 2D assumption in ptychography
\begin{equation}
\label{eq:conv_2d}
\tilde{\psi}_k(\vec{\xi}_{d,\bot}) = \iint \tilde{O}'_k(\vec{\xi}_{d,\bot} - \vec{\xi}_{i,\bot})\tilde{P'}(\vec{\xi}_{i,\bot})d^2\vec{\xi}_{i,\bot}.
\end{equation}
The 2D functions $O'$ and $P'$ correspond to ptychographic solutions one would expect to reconstruct for the sample of interest if a ptychographic solver is applied without taking into account any depth dependence. Our goal in this section will be to find out under which circumstances there exist functions $O'$ and $P'$ that produce an approximately equal scattered field to the field predicted by the true, three-dimensional $O$ and $P$. We can then attempt to find upper limits on the sample thickness for which these approximations hold.

The approach will be the following: First we consider the case of a single ``slab'' for which we will assume that its object function is separable between its lateral (2D, $\bot$) and axial (1D, $z$) dimensions. For this layer, we find functions $O'$ and $P'$ under various levels of approximation. Next, we make the problem more specific by defining two experimental geometries with finite numerical aperture; one for transmission and one for reflection-type experiments. This allows us to derive constraints on the maximal sample thickness for these specific geometries.

We start out by separating the lateral object function from the effect of depth by assuming the object is separable, i.e.
\begin{equation}
\label{eq:layered_slices}
\hat{O}(\vec{\xi}) = \tilde{O}_\bot(\vec{\xi}_\bot)\bar{O}_{z}(\xi_z).
\end{equation}
Note that the separability in the Fourier domain also means separability in real space. We find the predicted scattered field under the separable object by inserting Eq.~(\ref{eq:layered_slices}) into Eq.~(\ref{eq:FDT_impulse}):
\begin{equation}
\label{eq:conv3d_separable}
\hat{\psi}(\vec{\xi}_d) = \delta_{ES}(\vec{\xi}_d)\iiint \tilde{O}_\bot(\vec{\xi}_{d,\bot} - \vec{\xi}_{i,\bot})\bar{O}_{z}(\xi_{d,z} - \xi_{i,z})\hat{P}(\vec{\xi}_i)d^3\vec{\xi}_i.
\end{equation}
Working towards Eq.~(\ref{eq:conv_2d}), we inverse transform over $z$, and substitute $z=0$. Since we are now working towards a 2D description again, for conciseness we will only consider $\tilde{P}^+$; a full description would be a sum of the positive and negative contributions. The scattered field (coordinate $\vec{\xi}_d$) on the other hand will be taken at both the positive and negative half sphere, as our goal is to compare transmission and reflection geometries, which will have opposite signs. Performing the integral over $\xi_{d,z}$ from the inverse transform and over $\xi_{z,i}$ with the impulse functions in $\vec{\xi}_d$ and $\vec{\xi}_i$ allows us to substitute $\xi_{d.z}=\pm k_z(\vec{\xi}_{d,\bot})$ and $\xi_{i,z} = k_z(\vec{\xi}_{i,\bot})$ to give us
\begin{equation}
\label{eq:almost_conv2d}
\tilde{\psi}(\vec{\xi}_{d,\bot};0) = \iint \frac{\pm k_z(\vec{\xi}_{d,\bot})k_z(\vec{\xi}_{i,\bot})}{k^2}
    \tilde{O}_\bot(\vec{\xi}_{d,\bot} - \vec{\xi}_{i,\bot})
    \bar{O}_{z}\left(\pm k_z(\vec{\xi}_{d,\bot}) - k_z(\vec{\xi}_{i,\bot})\right)
    \tilde{P}^+(\vec{\xi}_{i,\bot})d^2\vec{\xi}_{i,\bot}.
\end{equation}
This expression is a 2D equivalent of Eq.~(\ref{eq:conv_2d}), however it is not yet in the required form, since the appearance of the individual $\vec{\xi}_{d,\bot}$ and $\vec{\xi}_{i,\bot}$ coordinates prevent writing the expression as a convolution. To rewrite this near-convolution towards the form of Eq.~(\ref{eq:conv_2d}), we must make certain assumptions on $\bar{O}_z$ and $k_z(\vec{\xi})$. We will now formulate some of these assumptions, which will later be used to derive conditions on the sample thickness.

The simplest assumption to make is to neglect the influence of $\bar{O}_z$ entirely;
\begin{equation}
\label{eq:Oz_strictest}
\bar{O}_z(\xi_z)\approx 1,
\end{equation}
for all $\xi_z$ in the integral. Although this way we ensure a true reconstruction, it also places the most stringent conditions on maximum thickness. Since nearly equivalent assumptions exist which are much less strict, this assumption is not very practical and we will disregard it from further analysis. A less strict assumption is produced by taking the coordinate functions at the central incoming and outgoing wavevectors $\xi_i=\vec{k}_{i,0}$ and $\xi_d=\vec{k}_{d,0}$. Then we can substitute $k_z(\vec{\xi}_{d,\bot}) \approx k_{z,d,0}$ and $k_z(\vec{\xi}_{i,\bot}) \approx k_{z,i,0}$ into Eq.~(\ref{eq:almost_conv2d}):
\begin{equation}
\label{eq:Oz_strict}
\frac{\pm k_z(\vec{\xi}_{d,\bot})k_z(\vec{\xi}_{i,\bot})}{k^2}
     \bar{O}_{z}\left(\pm k_z(\vec{\xi}_{d,\bot}) - k_z(\vec{\xi}_{i,\bot})\right)
     \approx 
     \frac{\pm k_{d,z,0} k_{i,z,0}}{k^2}
     \bar{O}_z(\pm k_{d,z,0}-k_{i,z,0}).
\end{equation}
This approximation does not lead to large changes in Eq.~(\ref{eq:almost_conv2d}) as long as $\bar{O}_z$ does not vary much over the entire domain of $k_z(\vec{\xi}_d)$ and $k_z(\vec{\xi}_i)$ in the integral, and the $k_z$ values do not vary by too much with respect to the central wavevector for both $\vec{\xi}_i$ and $\vec{\xi}_d$. Because the demands this places on the variation of $\bar{O}_z$ are still significant, this will be referred to as the ``strict'' approximation.

It is important to note that this condition controls the absolute variation of $\bar{O}_z$ over the sampled range, but does not by itself guarantee that the approximation in Eq.~(\ref{eq:Oz_strict}) remains informative. In particular, if the central scattering vector $\pm \vec{k}_{d,0} -\vec{k}_{i,0}$ lies near a zero of $\bar{O}_z$, then the zeroth-order term of the strict approximation becomes singular at the expansion point: the central contribution vanishes while off-axis contributions may remain finite. Such a situation may occur if the specular reflection geometry is such that it corresponds to a destructive Bragg condition, where complex contributions to the scattering of the central wavevectors average to precisely 0. In that regime, even a small absolute variation of $\bar{O}_z$ over the sampled domain can be large relative to the strict approximation itself, so higher-order terms in the variation of $\bar{O}_z$ are still relevant to the total intensity.

To make the assumptions even less demanding, we can take the central point in the angular distribution in just one of the two coordinates, either $\vec{\xi}_d$ or $\vec{\xi}_i$, but keep the dependence on the other coordinate. Then we get the following cases:
\begin{equation}
\frac{\pm k_z(\vec{\xi}_{d,\bot})k_z(\vec{\xi}_{i,\bot})}{k^2}
    \approx
    \frac{\pm k_{z,d,0}k_z(\vec{\xi}_{i,\bot})}{k^2}
    =f_1(\vec{\xi}_{i,\bot}),
\end{equation}
\begin{equation}
\label{eq:Oz_relaxed_input}
\bar{O}_{z}\left(\pm k_z(\vec{\xi}_{d,\bot}) - k_z(\vec{\xi}_{i,\bot})\right) 
    \approx
    \bar{O}_z(\pm k_{z,d,0}-k_z(\vec{\xi}_{i,\bot})) 
    = f_2(\vec{\xi}_{i,\bot}), 
    \;\text{ or}
\end{equation}
\begin{equation}
\frac{\pm k_z(\vec{\xi}_{d,\bot})k_z(\vec{\xi}_{i,\bot})}{k^2}
    \approx
    \frac{\pm k_z(\vec{q}_{\bot} + \vec{k}_{i,\bot,0})k_{i,z,0}}{k^2}
    =g_1(\vec{q}_{\bot}),
\end{equation}
\begin{equation}
\label{eq:Oz_relaxed_output}
\bar{O}_{z}\left(\pm k_z(\vec{q}_\bot+\vec{\xi}_{i,\bot}) - k_z(\vec{\xi}_{i,\bot})\right) 
    \approx 
    \bar{O}_z(\pm k_z(\vec{q}_{\bot} + \vec{k}_{i,\bot,0}) - k_{i,z,0}) 
    = g_2(\vec{q}_\bot),
\end{equation}
where in the latter we have changed coordinates from $\vec{\xi}_{d,\bot} \rightarrow \vec{q}_\bot + \vec{\xi}_{i,\bot}\approx \vec{q}_\bot + \vec{k}_{i,\bot,0}$ such that it can be directly included in $O'$ of Eq.~(\ref{eq:conv_2d}). Eq.~(\ref{eq:Oz_relaxed_input}) needs to hold only within the domain of the detector coordinate $\vec{\xi}_d$, while Eq.~(\ref{eq:Oz_relaxed_output}) has the same constraint, but for the illumination coordinate $\vec{\xi}_i$. Since these approximations place less strict requirements on $\bar{O}_z$, these will be known as the ``relaxed'' thin-sample approximations, limited by either the detector range or the illumination range. Using these approximations, we find for the 2D equivalents of Eq.~(\ref{eq:conv_2d}):
Strict:
\begin{equation}
    \tilde{O}'(\vec{\xi}_\bot) = 
    \tilde{O}_\bot(\vec{\xi}_\bot)\frac{\pm k_{d,z,0} k_{i,z,0}}{k^2} \bar{O}_z(\pm k_{d,z,0}- k_{i,z,0}),
\end{equation}
\begin{equation}
\label{eq:strict_approx_functions}
\tilde{P}'(\vec{\xi}_\bot) 
    = \tilde{P}^+(\vec{\xi}_\bot),
\end{equation}
relaxed (detector):
\begin{equation}
    \tilde{O}'(\vec{\xi}_\bot) = 
    \tilde{O}_\bot(\vec{\xi}_\bot),
\end{equation}
\begin{equation}
\label{eq:relaxed_input_approx_functions}
\tilde{P}'(\vec{\xi}_\bot) = 
    \tilde{P}^+(\vec{\xi}_\bot)\frac{\pm k_{z,d,0}k_z(\vec{\xi}_{i,\bot})}{k^2}\bar{O}_z(\pm k_{d,z,0}-k_z(\vec{\xi}_\bot)),
\end{equation}
relaxed (illumination):
\begin{equation}
    \tilde{O}'(\vec{\xi}_\bot) = 
    \tilde{O}_\bot(\vec{\xi}_\bot)\frac{\pm k_z(\vec{\xi}_{\bot})k_{i,z,0}}{k^2}\bar{O}_z(\pm k_z(\vec{\xi}_{\bot} + \vec{k}_{i,\bot,0}) - k_{i,z,0}),
\end{equation}
\begin{equation}
\label{eq:relaxed_output_approx_functions}
\tilde{P}'(\vec{\xi}_\bot)= \tilde{P}^+(\vec{\xi}_\bot).
\end{equation}
which are obtained by demanding equality between Eq.~(\ref{eq:almost_conv2d}) and Eq.~(\ref{eq:conv_2d}). Note that the factor $\bar{O}_z$ in Eq.~(\ref{eq:strict_approx_functions}) is a constant, so it may be exchanged between $P'$ and $O'$. Since this ambiguity always exists for ``blind'' ptychography where both $P$ and $O$ are reconstructed\citep{fannjiang2020}, it is irrelevant whether the factor is placed into $O'$ or $P'$.

So far we have formalized constraints on the variation of the specimen function $\bar{O}_z$ in reciprocal space. We will now translate this to constraints on the spatial extent of the sample (i.e. the single layer thickness) in real space. Let us choose a function that limits the z-dependence in real space, for example a rectangle function of thickness $t$, with Fourier transform $\bar{O}_{z}(\xi_z) = t \text{sinc }(t\xi_z)$.

We find the maximum thickness $t_{max}$ by enforcing the argument of the $\text{sinc}$ function to vary by much less than $2\pi$ within the domains required by either Eq.~(\ref{eq:Oz_strict}), Eq.~(\ref{eq:Oz_relaxed_input}), Eq.~(\ref{eq:Oz_relaxed_output}). If this is the case, their respective approximations are valid. This constraint leads to
\begin{equation}
\label{eq:t_strict}
t_{\text{strict}} \ll \frac{2\pi}{\Delta(\pm\xi_{d,z}-\xi_{i,z})},
\end{equation}
\begin{equation}
\label{eq:t_relaxed_detector}
t_{\text{relaxed,d}} \ll \frac{2\pi}{\Delta(\xi_{d,z})},
\end{equation}
\begin{equation}
\label{eq:t_relaxed_illumination}
t_{\text{relaxed,i}} \ll \frac{2\pi}{\Delta(\xi_{i,z})}.
\end{equation}
The $\Delta(\cdot)$ indicates the range of its argument, i.e. $\Delta(a) = \max(a) -\min(a)$.  The ranges of $\xi_{d,z}$ and $\xi_{i,z}$ have been schematically shown for the transmission and reflection geometries in

\begin{figure}[!htbp]
\centering
\includegraphics[width=1\linewidth]{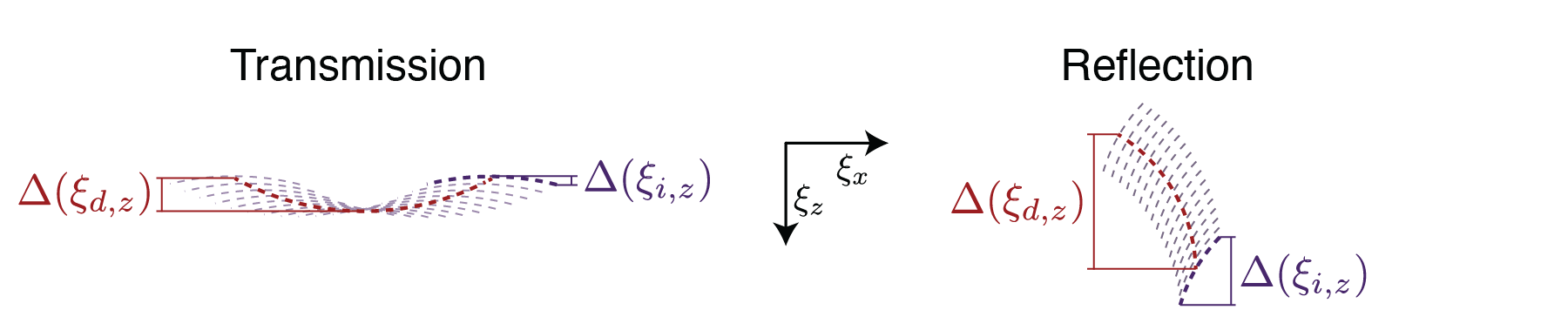}
\caption[]{The sampled $\xi_z$-ranges in the convolution of Eq.~(\ref{eq:FDT_impulse}) for the transmission geometry (left) and reflection geometry (right).}
\label{fig:placeholder}
\end{figure}

We see that the approximations made by Eq.~(\ref{eq:relaxed_input_approx_functions}), Eq.~(\ref{eq:relaxed_output_approx_functions}) differ only by swapping the distorting $\bar{O}_z$ between probe and object, depending on which approximations in Eq.~(\ref{eq:t_relaxed_detector}), Eq.~(\ref{eq:t_relaxed_illumination}) are satisfied. For brevity, we will summarize these two cases in terms of a ``limiting'' coordinate, whichever places the least strict requirement on the z-range assumed constant in $\bar{O}_z$:
\begin{equation}
\vec{\xi}_\text{lim} = \left\lbrace\begin{aligned}
    \vec{\xi}_d && \text{ if } \Delta(\xi_{d,z}) < \Delta(\xi_{i,z}), \\
    \vec{\xi}_i && \text{ if } \Delta(\xi_{d,z}) \geq\Delta(\xi_{i,z}),
\end{aligned}\right.
\end{equation}
and so
\begin{equation}
\label{eq:t_lim}
t_{\text{lim}} \ll \frac{2\pi}{\Delta(\xi_{\text{lim},z})}.
\end{equation}
Which coordinate is limiting determines where the distorting function $\bar{O}_z$ is expected to manifest itself. If $\vec{\xi}_{\text{lim}} = \vec{\xi}_d$, i.e.  the illumination has the larger $\xi_z$-range and the detector has the smaller $\xi_z$-range, then the error should manifest itself in the probe function $P'$ function via Eq.~(\ref{eq:relaxed_input_approx_functions}). Conversely, If $\vec{\xi}_{\text{lim}} = \vec{\xi}_i$, i.e. the detector has the larger $\xi_z$-range in Fourier space and the illumination has the smaller $\xi_z$-range, then the error will manifest itself in the object function $O'$ via Eq.~(\ref{eq:relaxed_output_approx_functions}).

\section{Reflection samples must be much thinner than transmission samples}

We have seen that there are several thin-sample approximations that can be made, depending on the exact range of $\xi_z$ values in the detector and illumination coordinates, which will be dictated by the experimental geometry. We can now apply the developed theory to predict the requirements of sample thickness on two different experimental geometries: reflection-mode and transmission-mode. We will show how the conditions from Eq.~(\ref{eq:t_strict}), Eq.~(\ref{eq:t_lim}) lead to vastly different requirements on sample thickness, where the reflection geometry may require orders of magnitude smaller thicknesses.

In the transmission geometry, the z-component of the Fourier coordinates only arises due to the curvature of the Ewald sphere. Approximating the curvature as a parabolic (paraxial approximation), we therefore obtain
\begin{equation}
\frac{t_{\text{strict}}}{\lambda}       \ll \frac{2}{\text{NA}_i^2+\text{NA}_d^2},
\end{equation}
\begin{equation}
\label{t_lim_trans}
\frac{t_{\text{lim}}}{\lambda}    \ll \frac{2}{\text{NA}_{\text{lim}}^2},
\end{equation}
where we define $\text{NA}_{\text{lim}} = \min(\text{NA}_{i}, \text{NA}_d)$ and we note that Eq.~(\ref{t_lim_trans}) has been previously derived in \citep{rodenburg1992}.

For the specular reflection geometry the story is quite different: now the appearance of $\xi_z$ in the diffraction integrals is dominated by the rotation of the Ewald sphere, and it is instead the curvature that may be neglected. This leads to
\begin{equation}
\frac{t_{\text{strict}}}{\lambda}       \ll \frac{1}{2\sin(\theta_i)(\text{NA}_i + \text{NA}_d)}
\end{equation}
\begin{equation}
\frac{t_{\text{lim}}}{\lambda}    \ll \frac{1}{2\sin(\theta_i) \;\text{NA}_{\text{lim}}}.
\end{equation}

Note that the negligence of the Ewald sphere curvature is only valid when the rotation is the dominating contribution to $\Delta(\xi_z)$. For the relaxed case, this is the case when $\theta_i \geq \frac{1}{4}\text{NA}_{\text{lim}}$ with $\theta_i$ in radians. There is also a reflection geometry which illuminates directly perpendicular to the sample plane and detect directly reflected light, i.e. where $\theta_i =0$, $\theta_d = \pi$. It is important to note that cases where $\theta_i = 0$, i.e. reflection ptychography under normal incidence, the limiting condition is again equivalent to the transmission case.

\begin{table}
\centering
\caption[]{The maximum thicknesses for an example geometry in EUV reflection ptychography.}
\label{tab:transmission_reflection_comparison}
\begin{tabular}{p{\dimexpr 0.333\linewidth-2\tabcolsep}p{\dimexpr 0.333\linewidth-2\tabcolsep}p{\dimexpr 0.333\linewidth-2\tabcolsep}}
\toprule
 & Transmission & Reflection \\
\hline
$t_{\text{strict}}$ & $\ll 22\lambda$ & $\ll 1.7\lambda$ \\
$t_{\text{lim}} = t_{\text{relaxed},i}$ & $\ll 800\lambda$ & $\ll 12\lambda$ \\
\bottomrule
\end{tabular}
\end{table}

To assess the implications of the differences between a transmission and a reflection geometry, let us take a numerical example corresponding to a realistic scenario in Extreme Ultraviolet (EUV) reflection ptychography: $\text{NA}_i = 0.05$, $\text{NA}_d = 0.3$, $\theta_i = \theta_d = 60^\circ$. This example has been worked out in Table~\ref{tab:transmission_reflection_comparison}. The decrease of a factor 10 in the strict case and a factor 70 in the relaxed case in the example highlights the generally much more stringent thin sample conditions for reflection geometries as compared to transmission geometries. In practical experiments the strict condition is rarely satisfied. The large relative difference between the transmission and reflection geometries deserves particular emphasis: evidently, reflection geometries are much more sensistive to depth than their transmission counterparts.

\section{Simulation verification}

To verify the predictions from the theory, our approach will be to first simulate diffraction patterns from a ground-truth model which fits the assumptions made previously, such that the predictions on the sample thickness and expected errors should hold. Next, we reconstruct these simulated datasets using a simple two-dimensional ptychographic solver. By analyzing the resulting $P'$ and $O'$ in the two-dimensional case we can verify our model predictions. Additionally, we can study whether the model that includes the effects of depth is able to correct the distortions in the 2D reconstructions and can again reconstruct the ground truth values.

A simple numerical model that fits the theory from the previous chapter is a two-layer reflection, where we are free to choose the lateral patterning, but we fix the depth function $\bar{O}_z$ to
\begin{equation}
O_z(z)          = \delta(z) + \delta(z-t),\\
\bar{O}_z(\xi_z)      \propto \cos(\frac{1}{2}t\xi_z),
\end{equation}
With $t$ again the thickness. For simplicity we will choose a binary lateral patterning for $\bar{O}_z$, similar to patterned nanostructures, corresponding to a Siemens star pattern for easy resolution assessment. We will refer to these reconstruction as the `2+1D'-case, since it includes both the lateral and axial dimensions, but the axial effects are handled differently from the lateral effects, i.e. they are not modeled by fully sampled arrays. This model is implemented numerically in real-space, by taking the exit field to be the sum of the contribution from the top layer and the bottom layer:
\begin{equation}
\label{eq:reflection_model_simplified_first}
\psi(x, y) = \psi_{t} + \psi_{b},
\end{equation}
where
\begin{equation}
\psi_{t} = O_\bot(x, y) P(x, y),
\end{equation}

\begin{equation}
\label{eq:reflection_model_simplified_last}
\psi_{b} = \mathcal{P}^{-}_{-t}\{O_\bot(x,y)\mathcal{P}^+_{t}\{P(x,y)\}\},
\end{equation}
where $\mathcal{P}^\pm_z$ is the off-axis angular spectrum propagator\citep{kim2015} in the $\pm z$-direction for a distance $z$. All simulations were run in dimensionless units of length proportional to the wavelength. The probe was chosen as an elongated circular aperture, spanning roughly half the real space field-of-view of a single diffraction pattern, to ensure that $\text{NA}_{\text{lim}}=\text{NA}_i$. Propagation to the detector was performed using a tilted forward model previously described in \citep{senhorst2024}. Reconstructions were performed on the simple two-dimensional datasets, with an initial probe conforming to the ground truth from the simulations, and an initial object set uniformly. Unless otherwise indicated, the probe was not included in the optimization, to keep the situation as close as possible to the theoretical description where we assume either $O$ or $P$ may receive the distortion due to $\bar{O}_z$.

\subsection{Reconstructed distortions in reflection ptychography}

Firstly we assess the effect of the distortions by $\bar{O}_z$ on the reconstructed lateral patterning for two-dimensional reflection ptychography experiments. Simulations and reconstructions were performed for an incident angle $\theta_i = 60^\circ$, such that $\cos(\theta_i)=0.5$. $t$ was varied in the range of $(0, 40\lambda)$ and $\text{NA}_d = 0.3$ was chosen for the detector $\text{NA}$.

The 2D reconstructions in reflection showed large, periodic variations in the residual loss after optimization with respect to the thickness, corresponding to a period of $1\lambda$ with the smallest residual loss at $t=n\lambda$ and the largest residual loss at $t=(n+0.5)\lambda$ for integer $n$. These values correspond exactly to thicknesses which satisfy the destructive (high residual loss) and constructive (low residual loss) Bragg interference conditions for the zero-order specular reflection. In our three dimensional model, this may be understood as values where $\bar{O}_z$ in Eq.~(\ref{eq:Oz_relaxed_output}) is either 0 (destructive) or maximal (constructive) for $\vec{\xi}_\bot=(0,0)$.

This also explains why elevated residual loss already appears at $t=0.5\lambda$, even though the thin-sample bounds derived above are still satisfied. At this thickness, the strict approximation samples $\bar{O}_z$ at the central incoming and outgoing directions, where the specular term is at a Bragg minimum and the corresponding zeroth-order contribution vanishes. The measured intensity is nevertheless not zero, because the finite illumination and detection apertures still sample nearby off-axis scattering vectors for which $\bar{O}_z$ remains non-zero. The resulting dataset is therefore not incompatible with the thin-sample conditions; rather, it is a case where the zeroth-order strict approximation becomes poorly conditioned and the variation of $\bar{O}_z$ across the aperture becomes the dominant contribution.

The total intensity of the diffraction patterns shows a similar variation to the reconstructed error, such that one might be tempted to attribute the higher residual loss to a lower signal-to-noise ratio (SNR). We note however that these simulations were performed assuming theoretical ideal conditions, without addition of noise or limitation of the dynamic range. This means that the signal to noise ratio is equal for all thicknesses, independent of the total intensity in the detected pattern, and therfore the distortions in the reconstructions are not caused by SNR variations.

\begin{figure}[!htbp]
\centering
\includegraphics[width=0.7\linewidth]{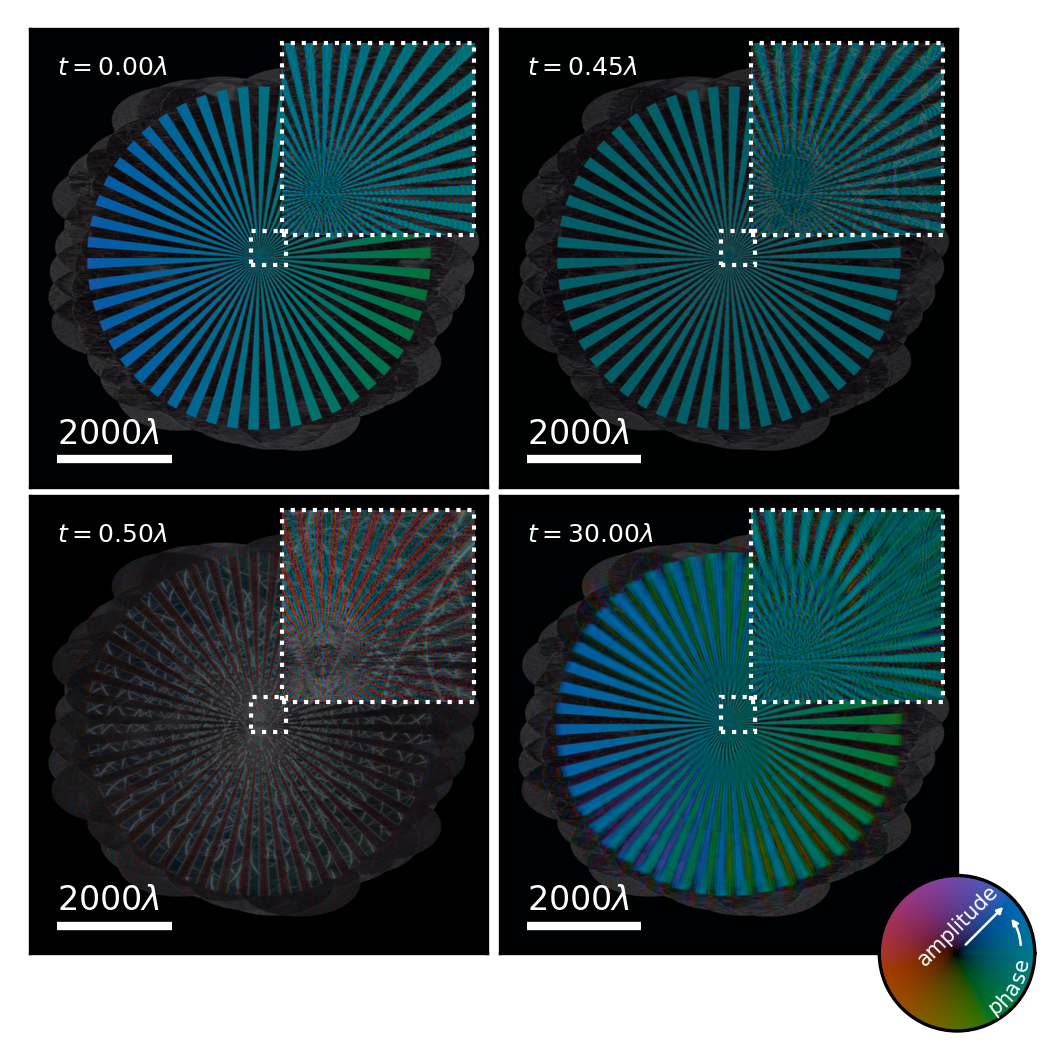}
\caption[]{The reconstructed two-dimensional function $O'$ for 3D simulated data, given a true thickness value of (from left to right) 0, $0.45\lambda$, a destructive Bragg condition at $0.5\lambda$ and a constructive Bragg condition at $30.0\lambda$.}
\label{fig:reconstruction_errors}
\end{figure}

The reconstructed $O'$ values hardly show any effect of the distortion by $\bar{O}_z$ for most low thickness values, with the notable exception when the thickness is very close to a destructive Bragg condition. In that case, the average of the object function $O'$ (corresponding to $\tilde{O}'(0,0)$) is nearly 0, with contrast only remaining near x-derivatives of the original object function $O$. This may be understood by linear approximation of the distortion function in Fourier space with respect to $\xi_x$ in Eq.~(\ref{eq:relaxed_output_approx_functions}): $\bar{O}_z(\xi_\bot)\approx a\xi_x$, resulting in $O'(\vec{r}) \propto \frac{d}{dx} O_\bot(\vec{r})$. When further increasing the thickness to $t$, the reconstructed $O'$ also exhibits clear errors for experiments in constructive Bragg conditions. These reconstructions have been shown in Fig.~\ref{fig:reconstruction_errors}.

\begin{figure}[!htbp]
\centering
\includegraphics[width=1\linewidth]{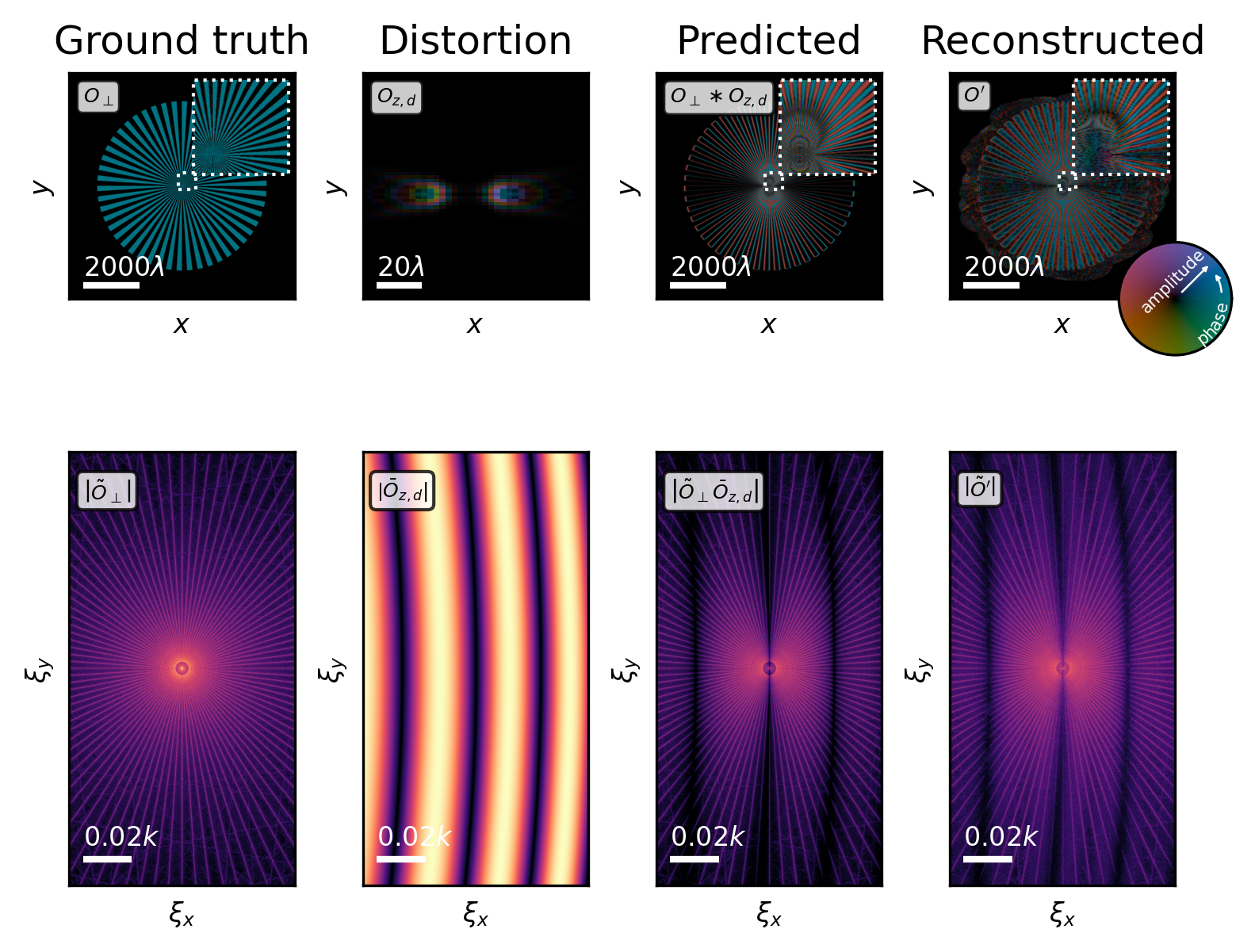}
\caption[]{A comparison between the real- and Fourier space representations of (from left to right), the ground truth lateral function $\tilde{O}_\bot(\vec{\xi})$, the approximation of the distorting function evaluated on the detection Ewald sphere $\bar{O}_{z,d}(\vec{\xi}) =\bar{O}_z(\pm k_z(\vec{\xi}_{\bot} + \vec{k}_{i,\bot,0}) - \vec{k}_{i,z,0})$, the ground truth after applying the predicted distortion $(\tilde{O} ~ \bar{O}_{z,d})$, and the reconstructed object $O'$. The simulation was based on a thickness of $20.5 \lambda$, corresponding to a Bragg minimum condition, as may be seen by the line of $|\bar{O}_{z.d}|$ which passes through the center of the Fourier space.}
\label{fig:oz_verification}
\end{figure}

To verify the prediction of the distorting function $\bar{O}_z$, the Fourier transformed version of the reconstructed object $\tilde{O}'$ was compared to the reconstruction predicted by the approximation theory in Eq.~(\ref{eq:relaxed_output_approx_functions}). The predicted reconstruction is computed by a multiplication in the Fourier domain between the lateral funtion $\tilde{O}_\bot$ and the approximated error $\bar{O}_z$, so the real space $O_z$ acts like a point-spread function on the true lateral object $O_\bot$. A comparison between the expected and the true reconstructions for a destructive Bragg thickness of $20.5\lambda$ shows excellent agreement between the prediction of the distortion and the reconstructed two-dimensional function, as can be seen in in Fig.~\ref{fig:oz_verification}.

\subsection{Optimization of the 2+1D model}

We have seen that two-dimensional reconstructions are not guaranteed to be sufficient for reconstructing weak-scattering samples, starting at thicknesses already corresponding to the first Bragg minimum. This raises the question if optimization of the 2+1D model will be able to correct the distortions that are clear in the two-dimensional case. Furthermore, it should be possible to optimize for the thickness if the initial guess is incorrect, since the  2+1D model is a function of the thickness $t$. To this end, another simulation was conducted to study optimization of the 2+1D model. This simulation was designed with the intention of accurately portraying realistic experimental conditions, instead of theoretically perfect ones. The probe was chosen based on a previous reconstruction of experimental data and Poisson noise was added to the detected diffraction patterns. The Poisson noise was based on a total illumination power of $5 ~\mu$W at an EUV wavelength of 18.5 nm with a 100 ms exposure time, for an average photon count per pixel of $11.7\cdot10^4$ for a 1980x1980-pixel detector in the case of perfect reflection of the probe beam. The detected diffraction patterns were quantized to the range of a 14-bit detector. The initial guess for the object function was set to unity.

\begin{figure}[!htbp]
\centering
\includegraphics[width=1\linewidth]{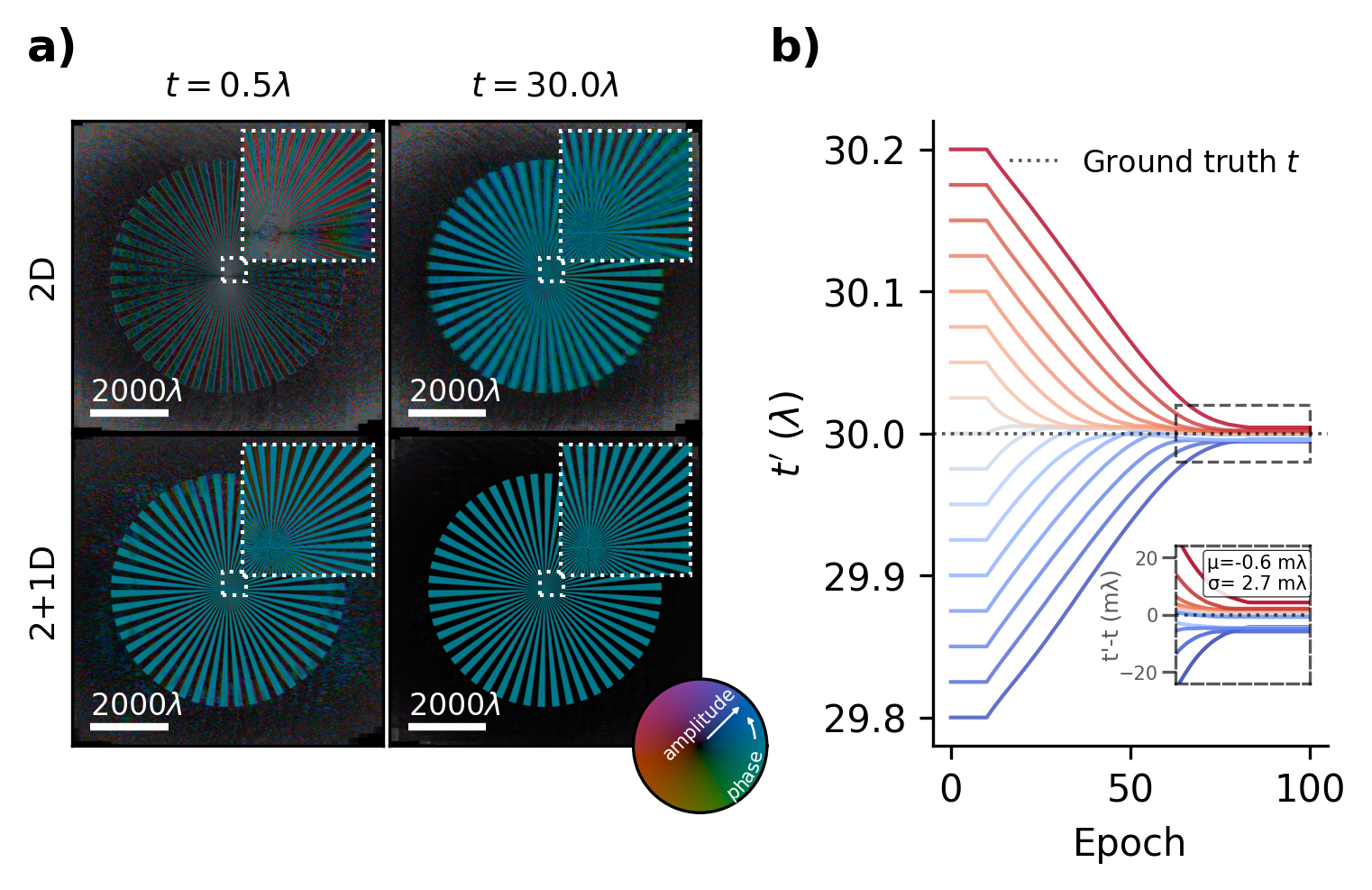}
\caption[]{a: The reconstructions of the two-dimensional model (top) versus the 2+1-dimensional model (bottom) for simulated data based on thicknesses of $0.5\lambda$ (left) and $30\lambda$ (right).  b: The optimization progression of the reconstructed thickness parameter $t'$ for the case of a true thickness $t$ of $30\lambda$, for varying initial guesses of $t'$. The inset is a y-scaled version of the full plot, indicating the mean from the ground truth $\mu$ and the standard deviation $\sigma$ away from the ground truth value in the final 10 epochs.}
\label{fig:3d_reconstructions}
\end{figure}

Firstly, we compare the case of the 2D to the 3D reconstructions in the Bragg destructive case, where the 2D reconstructions showed the most artifacts. As can be seen in Fig.~\ref{fig:3d_reconstructions}a, the errors in the lateral function that arise due to the destructive Bragg condition can be largely avoided when the 2+1D model is solved for, only parts in the edges of the field-of-view, where the object is illuminated by only a few scanning positions, still show some low-frequency deviations.

Another simulation was performed at the large ground-truth thickness of 30 $\lambda$, where the 2D models of the last sections also led to reconstruction errors near the higher spatial frequencies. Here, the initial guess for the reconstructed thickness $t'$ was offset from the ground truth value of $t$. Optimization of the thickness was enabled after 10 epochs to ensure stable reconstructions, while other reconstruction parameters were identical to the previous section. If the initial thickness was offset by more than roughly $0.2 \lambda$, optimization of the thickness would either stagnate or converge to a local minimum. The local minima seemingly appear for thicknesses at positions displaced by roughly a full wavelength from the true thickness. This may be explained by noting that for these thicknesses, the optimization most closely fits the Bragg minima and maxima to the nearest thickness where they more or less align with the data. For simulations with initial guesses inside of this domain of convergence, the thickness could reliably converge to the true value, as can be seen in Fig.~\ref{fig:3d_reconstructions}b. The final accuracy of the optimized thicknesses, measured as the root mean square deviation from the ground truth, was $\pm 2.7 ~$m$\lambda$, or 50 pm for the simulation wavelength of 18.5 nm.

\section{Discussion}

The results as presented above clearly indicate that the developed theory may be used to predict assess the validity of the two-dimensional thin sample approximation in reflection ptychography, however it still relies on the assumptions of weak scattering and separability. For samples which satisfy these assumptions, we have derived practical limits on the sample thickness for which a 2D approximation is still valid. The theory additionally predicts the appearance of distortions in 2D reconstructions whenever the strictest thickness condition is not satisfied, thus extending the original theory of ptychography\citep{rodenburg1992} to reflection geometries and making explicit the influence of thickness on its reconstructions. We confirmed the predicted distortions appear in simulation, thereby validating the predicted thickness limits. Finally, we demonstrate that inclusion of the depth in the forward model can lead to reconstructions without distortions, and can additionally optimize for the thickness for an incorrect initial guess.

Plenty of research has been published which demonstrated experimentally that ptychography can be applied in a reflection geometry. This is not in contradiction with our research results reported here, where we also show that there are many thicknesses for which simple 2D models can provide reconstructions which are nearly indistinguishable from their ground-truth values; ptychography is evidently fairly robust with respect to noise sources and specific errors that do not fit into the original modeling. However, our present work shows that there is always a possibility of an unknown modeling inaccuracy giving rise to unexpected artifacts, especially when those models build on theory for which it is not validated. As a more robust theory for reflection ptychography over extended depth ranges is currently not available, one should therefore always beware of the possibility that reconstructed features of interest are not physical, but instead arise from modeling inconsistencies.

A remaining question is whether the assumptions of weak scattering and separability made in this work are also valid in experimental settings, where reflection ptychography is intended to be usedl; this work does not provide an answer to this question. One should note however that we have only extended the theoretical framework that forms the basis of ptychography. In other words, if the theory as presented in this work is invalid for experimental use because experiments of practical interest do not satisfy its assumptions, then there is currently no available theory that is valid in describing the outcomes of such experiments. This work may be considered a first step towards the development of a theory of reflection ptychography that accurately models the inherent three-dimensional nature of the reflected field of samples which are more than several wavelengths thick.

\section{Conclusion}

This work demonstrates that application of the theory developed for transmission ptychography by performing two-dimensional reconstructions in reflection geometries at non-normal incident angles is not guaranteed to provide accurate reconstructions. If the measurement geometry and sample thickness are known, the distortion may be predicted using the developed theory. The distortion may be small for low thicknesses and away from Bragg minima, but whenever Bragg minima appear in the field-of-view of the detector, artifacts are bound to appear. It may take only a small change in modeling, namely inclusion of the correct depth effects in the propagation model, to prevent the distortions from appearing in the reconstruction outcome and thus to increase the range of sample thicknesses for which reflection ptychography may be safely applied. Whether or not the depth effects used in the simulations in this work are experimentally valid, their inclusion in the modeling gives the additional benefit of allowing one to reconstruct the thickness of samples of interest. This hints at the potential use of reflection ptychography as a means for imaging depth-resolved samples directly from reflection-ptychographic data, without need for variation of the angle-of-incidence or spectral diversity; an exciting prospect that warrants further study.

\begin{backmatter}

\bmsection{Funding}

TKI Holland High Tech (TKI HTSM/22.0220)

\bmsection{Disclosures}
WC: ASML Netherlands B.V. (E).

\bmsection{Data Availability Statement}
Data underlying the results are available upon resonable request.

\end{backmatter}

\bibliography{main.bib}

@inproceedings{bogdanowicz2025,
	note = {[Online; accessed 2025-05-23]},
	author = {Bogdanowicz, J. and Charley, A.-L. and Leray, P. and Liu, R. G.},
	booktitle = {Metrology, {Inspection}, and {Process} {Control} {XXXIX}},
	doi = {10.1117/12.3052399},
	year = {2025},
	month = {4},
	pages = {8--15},
	organization = {SPIE},
	title = {3D {Metrology} and {Inspection} to {Enable} the {Rise} of {Stacked} {Transistors}, {Wafers}, and {Chips}},
	url = {\url{https://www.spiedigitallibrary.org/conference-proceedings-of-spie/13426/1342604/3D-metrology-and-inspection-to-enable-the-rise-of-stacked/10.1117/12.3052399.full}},
	volume = {13426},
}

@article{gardner2012,
	note = {[Online; accessed 2023-03-29]},
	author = {Gardner, Dennis F. and Zhang, Bosheng and Seaberg, Matthew D. and Martin, Leigh S. and Adams, Daniel E. and Salmassi, Farhad and Gullikson, Eric and Kapteyn, Henry and Murnane, Margaret},
	journal = {Optics Express},
	doi = {10.1364/OE.20.019050},
	issn = {1094-4087},
	number = {17},
	year = {2012},
	month = {8},
	pages = {19050},
	title = {High {Numerical} {Aperture} {Reflection} {Mode} {Coherent} {Diffraction} {Microscopy} {Using} {Off}-{Axis} {Apertured} {Illumination}},
	url = {\url{https://opg.optica.org/oe/abstract.cfm?uri=oe-20-17-19050}},
	howpublished = {\url{https://opg.optica.org/oe/abstract.cfm?uri=oe-20-17-19050}},
	volume = {20},
}

@article{seaberg2014,
	author = {Seaberg, Matthew D. and Zhang, Bosheng and Gardner, Dennis F. and Shanblatt, Elisabeth R. and Murnane, Margaret M. and Kapteyn, Henry C. and Adams, Daniel E.},
	journal = {Optica, Vol. 1, Issue 1, pp. 39-44},
	doi = {10.1364/OPTICA.1.000039},
	number = {1},
	year = {2014},
	month = {7},
	pages = {39--44},
	publisher = {Optica Publishing Group},
	title = {Tabletop {Nanometer} {Extreme} {Ultraviolet} {Imaging} in an {Extended} {Reflection} {Mode} {Using} {Coherent} {Fresnel} {Ptychography}},
	url = {\url{https://opg.optica.org/viewmedia.cfm?uri=optica-1-1-39&seq=0&html=true}},
	volume = {1},
}

@article{rodenburg1992,
	note = {[Online; accessed 2023-03-29]},
	author = {Rodenburg, J. M. and Bates, R. H. T.},
	journal = {Philosophical Transactions: Physical Sciences and Engineering},
	issn = {0962-8428},
	number = {1655},
	year = {1992},
	pages = {521--553},
	publisher = {The Royal Society},
	title = {The {Theory} of {Super}-{Resolution} {Electron} {Microscopy} {Via} {Wigner}-{Distribution} {Deconvolution}},
	url = {\url{https://www.jstor.org/stable/53996}},
	volume = {339},
}

@article{odstrcil2015,
	note = {[Online; accessed 2026-03-09]},
	author = {Odstrcil, M. and Bussmann, J. and Rudolf, D. and Bresenitz, R. and Miao, Jianwei and Brocklesby, W. S. and Juschkin, L.},
	journal = {Optics Letters},
	doi = {10.1364/OL.40.005574},
	issn = {1539-4794},
	number = {23},
	year = {2015},
	month = {12},
	pages = {5574--5577},
	publisher = {Optica Publishing Group},
	title = {Ptychographic {Imaging} with a {Compact} {Gas}--{Discharge} {Plasma} {Extreme} {Ultraviolet} {Light} {Source}},
	url = {\url{https://opg.optica.org/ol/abstract.cfm?uri=ol-40-23-5574}},
	volume = {40},
}

@article{zhang2016,
	author = {Zhang, Bosheng and Gardner, Dennis F and Seaberg, Matthew H and Shanblatt, Elisabeth R and Porter, Christina L and Karl, Robert JR and Mancuso, Christopher A and Kapteyn, Henry C and Murnane, Margaret M and Adams, Daniel E and Wilson, D and Rudolf, D and Weier, C and Adam, R and Winkler, G and Fr{\" o}mter, R and Danylyuk, S and Bergmann, K and Gr{\" u}tzmacher, D and Schneider, C M and Juschkin, L and Strachan, J P and Medeiros-Ribeiro, G and Yang, J J and Zhang, M-x and Miao, F and Goldfarb, I and Holt, M and Rose, V and Williams, R S and Ade, H and Kirz, J and Hulbert, S L and Johnson, E D and Anderson, E and Kern, D},
	journal = {Optics Express, Vol. 24, Issue 16, pp. 18745-18754},
	doi = {10.1364/OE.24.018745},
	number = {16},
	year = {2016},
	month = {8},
	pages = {18745--18754},
	publisher = {Optica Publishing Group},
	title = {Ptychographic {Hyperspectral} {Spectromicroscopy} with an {Extreme} {Ultraviolet} {High} {Harmonic} {Comb}},
	url = {\url{https://opg.optica.org/viewmedia.cfm?uri=oe-24-16-18745&seq=0&html=true}},
	volume = {24},
}

@article{odstrcil2016,
	note = {[Online; accessed 2026-03-09]},
	author = {Odstrcil, M. and Baksh, P. and Boden, S. A. and Card, R. and Chad, J. E. and Frey, J. G. and Brocklesby, W. S.},
	journal = {Optics Express},
	doi = {10.1364/OE.24.008360},
	issn = {1094-4087},
	number = {8},
	year = {2016},
	month = {4},
	pages = {8360--8369},
	publisher = {Optica Publishing Group},
	title = {Ptychographic {Coherent} {Diffractive} {Imaging} with {Orthogonal} {Probe} {Relaxation}},
	url = {\url{https://opg.optica.org/oe/abstract.cfm?uri=oe-24-8-8360}},
	volume = {24},
}

@article{shanblatt2016,
	note = {[Online; accessed 2023-03-29]},
	author = {Shanblatt, Elisabeth R. and Porter, Christina L. and Gardner, Dennis F. and Mancini, Giulia F. and Karl, Robert M. Jr. and Tanksalvala, Michael D. and Bevis, Charles S. and Vartanian, Victor H. and Kapteyn, Henry C. and Adams, Daniel E. and Murnane, Margaret M.},
	journal = {Nano Letters},
	doi = {10.1021/acs.nanolett.6b01864},
	issn = {1530-6984},
	number = {9},
	year = {2016},
	month = {9},
	pages = {5444--5450},
	publisher = {American Chemical Society},
	title = {Quantitative {Chemically} {Specific} {Coherent} {Diffractive} {Imaging} of {Reactions} at {Buried} {Interfaces} with {Few} {Nanometer} {Precision}},
	url = {\url{https://doi.org/10.1021/acs.nanolett.6b01864}},
	volume = {16},
}

@article{porter2017,
	note = {[Online; accessed 2023-03-29]},
	author = {Porter, Christina L. and Tanksalvala, Michael and Gerrity, Michael and Miley, Galen and Zhang, Xiaoshi and Bevis, Charles and Shanblatt, Elisabeth and Karl, Robert and Murnane, Margaret M. and Adams, Daniel E. and Kapteyn, Henry C.},
	journal = {Optica},
	doi = {10.1364/OPTICA.4.001552},
	issn = {2334-2536},
	number = {12},
	year = {2017},
	month = {12},
	pages = {1552},
	title = {General-{Purpose}, {Wide} {Field}-of-{View} {Reflection} {Imaging} with a {Tabletop} 13 {Nm} {Light} {Source}},
	url = {\url{https://opg.optica.org/abstract.cfm?URI=optica-4-12-1552}},
	howpublished = {\url{https://opg.optica.org/abstract.cfm?URI=optica-4-12-1552}},
	volume = {4},
}

@article{karl2018,
	author = {Karl, Robert M. and Mancini, Giulia F. and Knobloch, Joshua L. and Frazer, Travis D. and Hernandez-Charpak, Jorge N. and Abad, Bego{\~ n}a and Gardner, Dennis F. and Shanblatt, Elisabeth R. and Tanksalvala, Michael and Porter, Christina L. and Bevis, Charles S. and Adams, Daniel E. and Kapteyn, Henry C. and Murnane, Margaret M.},
	journal = {Science Advances},
	doi = {10.1126/SCIADV.AAU4295},
	number = {10},
	year = {2018},
	month = {10},
	pages = {4295--4295},
	publisher = {American Association for the Advancement of Science},
	title = {Full-{Field} {Imaging} of {Thermal} and {Acoustic} {Dynamics} in an {Individual} {Nanostructure} {Using} {Tabletop} {High} {Harmonic} {Beams}},
	url = {\url{https://www.science.org}},
	volume = {4},
}

@article{tanksalvala2021,
	author = {Tanksalvala, Michael and Porter, Christina L. and Esashi, Yuka and Wang, Bin and Jenkins, Nicholas W. and Zhang, Zhe and Miley, Galen P. and Knobloch, Joshua L. and McBennett, Brendan and Horiguchi, Naoto and Yazdi, Sadegh and Zhou, Jihan and Jacobs, Matthew N. and Bevis, Charles S. and Karl, Robert M. and Johnsen, Peter and Ren, David and Waller, Laura and Adams, Daniel E. and Cousin, Seth L. and Liao, Chen Ting and Miao, Jianwei and Gerrity, Michael and Kapteyn, Henry C. and Murnane, Margaret M.},
	journal = {Science Advances},
	doi = {10.1126/SCIADV.ABD9667},
	number = {5},
	year = {2021},
	month = {1},
	publisher = {American Association for the Advancement of Science},
	title = {Nondestructive, {High}-{Resolution}, {Chemically} {Specific} 3D {Nanostructure} {Characterization} {Using} {Phase}-{Sensitive} {EUV} {Imaging} {Reflectometry}},
	url = {/pmc/articles/PMC7840142/},
	volume = {7},
}

@article{esashi2023,
	note = {[Online; accessed 2023-12-22]},
	author = {Esashi, Yuka and Jenkins, Nicholas W. and Shao, Yunzhe and Shaw, Justin M. and Park, Seungbeom and Murnane, Margaret M. and Kapteyn, Henry C. and Tanksalvala, Michael},
	journal = {Review of Scientific Instruments},
	doi = {10.1063/5.0175860},
	issn = {0034-6748},
	number = {12},
	year = {2023},
	month = {12},
	pages = {123705},
	publisher = {AIP Publishing},
	title = {Tabletop {Extreme} {Ultraviolet} {Reflectometer} for {Quantitative} {Nanoscale} {Reflectometry}, {Scatterometry}, and {Imaging}},
	url = {\url{https://doi.org/10.1063/5.0175860}},
	volume = {94},
}

@article{lu2023,
	note = {[Online; accessed 2024-03-05]},
	author = {Lu, Haoyan and Odstr{\v c}il, Michal and Pooley, Charles and Biller, Jan and Mebonia, Mikheil and He, Guanze and Praeger, Matthew and Juschkin, Larissa and Frey, Jeremy and Brocklesby, William},
	journal = {Ultramicroscopy},
	doi = {10.1016/j.ultramic.2023.113720},
	issn = {0304-3991},
	year = {2023},
	month = {7},
	pages = {113720},
	title = {Characterisation of {Engineered} {Defects} in {Extreme} {Ultraviolet} {Mirror} {Substrates} {Using} {Lab}-{Scale} {Extreme} {Ultraviolet} {Reflection} {Ptychography}},
	url = {\url{https://www.sciencedirect.com/science/article/pii/S0304399123000372}},
	howpublished = {\url{https://www.sciencedirect.com/science/article/pii/S0304399123000372}},
	volume = {249},
}

@article{shao2024,
	note = {[Online; accessed 2024-09-09]},
	author = {Shao, Yifeng and Weerdenburg, Sven and Seifert, Jacob and Urbach, H. Paul and Mosk, Allard P. and Coene, Wim},
	journal = {Light: Science \& Applications},
	doi = {10.1038/s41377-024-01558-3},
	issn = {2047-7538},
	number = {1},
	year = {2024},
	month = {8},
	pages = {196},
	publisher = {Nature Publishing Group},
	title = {Wavelength-{Multiplexed} {Multi}-{Mode} {EUV} {Reflection} {Ptychography} {Based} on {Automatic} {Differentiation}},
	url = {\url{https://www.nature.com/articles/s41377-024-01558-3}},
	volume = {13},
}

@article{tanksalvala2024,
	note = {[Online; accessed 2024-11-01]},
	author = {Tanksalvala, Michael and Kos, Anthony and Wisser, Jacob and Diddams, Scott and Nembach, Hans T. and Shaw, Justin M.},
	journal = {Physical Review Applied},
	doi = {10.1103/PhysRevApplied.21.064047},
	number = {6},
	year = {2024},
	month = {6},
	pages = {064047},
	publisher = {American Physical Society},
	title = {Element-{Specific} {High}-{Bandwidth} {Ferromagnetic} {Resonance} {Spectroscopy} with a {Coherent} {Extreme}-{Ultraviolet} {Source}},
	url = {\url{https://link.aps.org/doi/10.1103/PhysRevApplied.21.064047}},
	volume = {21},
}

@article{senhorst2024,
	author = {Senhorst, Sander and Shao, Yifeng and Weerdenburg, Sven and Horsten, Roland and Porter, Christina and Coene, Wim},
	journal = {Optics Express},
	doi = {10.1364/OE.542569},
	number = {25},
	year = {2024},
	month = {12},
	pages = {44017--44030},
	publisher = {Optica Publishing Group},
	title = {Mitigating {Tilt}-{Induced} {Artifacts} in {Reflection} {Ptychography} via {Optimization} of the {Tilt} {Angles}},
	url = {\url{https://opg.optica.org/oe/abstract.cfm?URI=oe-32-25-44017}},
	volume = {32},
}

@article{gao2025,
	note = {[Online; accessed 2025-12-09]},
	author = {Gao, Yun and You, Qijun and Lu, Peixiang and Cao, Wei},
	journal = {Optics and Lasers in Engineering},
	doi = {10.1016/j.optlaseng.2025.109076},
	issn = {0143-8166},
	year = {2025},
	month = {10},
	pages = {109076},
	title = {Interface and {Spectrum} {Multiplexing} {Ptychographic} {Reflection} {Microscopy}},
	url = {\url{https://www.sciencedirect.com/science/article/pii/S0143816625002623}},
	howpublished = {\url{https://www.sciencedirect.com/science/article/pii/S0143816625002623}},
	volume = {193},
}

@book{goodman1996,
	author = {Goodman, Joseph W.},
	isbn = {978-0-07-024254-8},
	year = {1996},
	publisher = {McGraw-Hill},
	title = {Introduction to {Fourier} {Optics}},
}

@article{onural2006,
	author = {Onural, Levent},
	journal = {Journal of Mathematical Analysis and Applications},
	doi = {10.1016/J.JMAA.2005.07.012},
	number = {1},
	year = {2006},
	month = {10},
	pages = {18--27},
	publisher = {Academic Press},
	title = {Impulse {Functions} over {Curves} and {Surfaces} and {Their} {Applications} to {Diffraction}},
	volume = {322},
}

@article{wolf1969,
	author = {Wolf, Emil},
	journal = {Optics Communications},
	doi = {10.1016/0030-4018(69)90052-2},
	number = {4},
	year = {1969},
	month = {9},
	pages = {153--156},
	publisher = {North-Holland},
	title = {Three-{Dimensional} {Structure} {Determination} of {Semi}-{Transparent} {Objects} from {Holographic} {Data}},
	volume = {1},
}

@article{zhou2021,
	author = {Zhou, Kevin C. and Qian, Ruobing and Dhalla, Al-Hafeez and Farsiu, Sina and Farsiu, Sina and Izatt, Joseph A. and Izatt, Joseph A.},
	journal = {Advances in Optics and Photonics, Vol. 13, Issue 2, pp. 462-514},
	doi = {10.1364/AOP.417102},
	number = {2},
	year = {2021},
	month = {6},
	pages = {462--514},
	publisher = {Optica Publishing Group},
	title = {Unified {K}-{Space} {Theory} of {Optical} {Coherence} {Tomography}},
	url = {\url{https://opg.optica.org/viewmedia.cfm?uri=aop-13-2-462&seq=0&html=true}},
	volume = {13},
}

@article{fannjiang2020,
	note = {[Online; accessed 2023-06-02]},
	author = {Fannjiang, Albert and Chen, Pengwen},
	journal = {Inverse Problems},
	doi = {10.1088/1361-6420/ab6504},
	issn = {0266-5611},
	number = {4},
	year = {2020},
	month = {2},
	pages = {045005},
	publisher = {IOP Publishing},
	title = {Blind {Ptychography}: Uniqueness and {Ambiguities}},
	url = {\url{https://dx.doi.org/10.1088/1361-6420/ab6504}},
	volume = {36},
}

@article{kim2015,
	note = {[Online; accessed 2025-12-09]},
	author = {Kim, Yong-Hae and Byun, Chun-Won and Oh, Himchan and Pi, Jae-Eun and Choi, Ji-Hun and Kim, Gi Heon and Lee, Myung-Lae and Ryu, Hojun and Hwang, Chi-Sun},
	journal = {Optics Communications},
	doi = {10.1016/j.optcom.2015.03.013},
	issn = {0030-4018},
	year = {2015},
	month = {8},
	pages = {31--37},
	title = {Off-{Axis} {Angular} {Spectrum} {Method} with {Variable} {Sampling} {Interval}},
	url = {\url{https://www.sciencedirect.com/science/article/pii/S0030401815002072}},
	howpublished = {\url{https://www.sciencedirect.com/science/article/pii/S0030401815002072}},
	volume = {348},
}






\end{document}